\documentclass[runningheads]{llncs}
\usepackage{graphicx}
\begin{document}
\title{Evaluation of the Programming Skills of Large Language Models} 
%
%
\author{Luc Bryan Heitz\inst{1}\orcidID{0009-0000-9279-2751} \and
Joun Chamas\inst{1} \and
Christopher Scherb\inst{1}\orcidID{0000-0001-6116-5093}}
\authorrunning{L. Heitz et al.}
%
\institute{University of Applied Sciences and Arts, Northwestern Switzerland \\
\email{luc.heitz@fhnw.ch, joun.chamas@students.fhnw.ch, christopher.scherb@fhnw.ch}\\
}
\maketitle              
\begin{abstract}
The advent of Large Language Models (LLM) has revolutionized the efficiency and speed with which tasks are completed, marking a significant leap in productivity through technological innovation. As these chatbots tackle increasingly complex tasks, the challenge of assessing the quality of their outputs has become paramount. This paper critically examines the output quality of two leading LLMs, OpenAI's ChatGPT and Google's Gemini AI, by comparing the quality of programming code generated in both their free versions. Through the lens of a real-world example coupled with a systematic dataset, we investigate the code quality produced by these LLMs. Given their notable proficiency in code generation, this aspect of chatbot capability presents a particularly compelling area for analysis. Furthermore, the complexity of programming code often escalates to levels where its verification becomes a formidable task, underscoring the importance of our study. This research aims to shed light on the efficacy and reliability of LLMs in generating high-quality programming code, an endeavor that has significant implications for the field of software development and beyond.

\end{abstract}
\section{Introduction}
The integration of Artificial Intelligence (AI) into the realm of software creation marks a pivotal shift in the methodologies and practices of programming, ushering in a new era of efficiency and inclusivity. AI chatbots, with their rapid code generation capabilities, stand at the forefront of this transformation, promising to redefine the traditional paradigms of software development. The allure of these technologies lies not only in their speed but in their potential to democratize coding, making it accessible to a wider audience with varied skill levels. However, the advent of AI in programming also introduces a complex web of challenges, primarily concerning the quality, safety, and maintenance of the generated code. These issues are of paramount importance, as inaccuracies or flaws in the code could precipitate severe consequences, including security vulnerabilities and operational failures.

Acknowledging these critical considerations, our investigation meticulously evaluates the code produced by two of the industry's frontrunners in AI chatbot technology: OpenAI's ChatGPT and Google's Gemini AI. Through the adoption of standardized testing frameworks and scenarios, our analysis endeavors to ensure an equitable and comprehensive assessment across a spectrum of programming challenges. This structured approach is designed to not only gauge the performance of these AI chatbots under varied conditions but also to mirror the intricate and multifaceted nature of real-world software development projects.

Our exploration extends beyond mere theoretical analysis; we venture into practical application by utilizing the AI-generated code to engineer a fully functional application. This pivotal phase of our study serves to empirically validate the operational efficacy of the generated code, scrutinizing its practicality, organization, and potential deficiencies, including the presence of code smells—a term used to describe indicators of deeper issues within the code's structure.

By subjecting AI-generated code to this dual-layered examination—both theoretical and empirical—our research aspires to illuminate the capabilities and limitations of Large Language Models in contemporary software development. Our objective is to underscore the criticality of meticulous quality assessment of AI-generated code, a factor that is indispensable for the realization of robust, secure, and efficient software solutions. This investigation not only seeks to enrich our comprehension of AI's role in software engineering but also to delineate the necessary protocols for testing and quality assurance that can maximize the benefits of AI technologies while mitigating their risks. Through this study, we aim to contribute to the broader discourse on leveraging AI in software development, paving the way for advancements that enhance the utility and reliability of AI-generated code in shaping the future of programming.

\section{Methodology}

This study aims to systematically evaluate and compare the performance of two leading AI chatbots, OpenAI's ChatGPT and Google's Gemini AI, in generating programming code. The core of our methodology revolves around the use of standardized datasets and prompts to ensure that our evaluation is both reproducible and reflective of each chatbot's capabilities in a variety of coding tasks. By focusing on reproducibility, we address the inherent variability in AI-generated responses, ensuring that our findings are robust and reliable.

\subsection{ Datasets}

We selected two widely recognized datasets for our evaluation: HumanEval and ClassEval. These datasets are specifically designed for assessing code generation capabilities and contain a series of coding tasks along with corresponding prompts that are used to instruct the AI chatbots. Each dataset offers a diverse set of challenges that are representative of real-world programming scenarios, making them ideal for our analysis.

 \begin{itemize}
     \item  HumanEval: This dataset comprises a range of programming exercises that require understanding and implementing algorithmic solutions. It is designed to test the chatbots' ability to generate functional code based on a given problem statement.
     \item ClassEval: Similar to HumanEval, ClassEval focuses on object-oriented programming tasks. It challenges the chatbots to create class structures, methods, and properties based on specific requirements, thereby assessing their understanding of more complex programming concepts.
 \end{itemize}

\subsection{Data Collection Process}

The data collection process involves the following steps:

\begin{enumerate}
    \item Query: Each task in the HumanEval and ClassEval datasets, come with an standardized prompt that clearly outlines the coding challenge. These prompts are designed to be as specific as possible to minimize ambiguity and ensure the tasks are understandable.
    \item Execution of Queries: Each prompt is submitted to both OpenAI's ChatGPT and Google's Gemini AI multiple times. For ChatGPT we use the API. This repetition is crucial for capturing the variability in responses, as AI chatbots may generate different solutions for the same prompt on different occasions.
    \item Result Aggregation: The responses from each submission are collected and analyzed. Given the variability of AI-generated code, we average the results of the multiple submissions to obtain a representative performance metric for each chatbot on each task.
\end{enumerate}

\subsection{Analysis and Discussion}
We critically analyze the aggregated and visualized data to draw conclusions about the overall capabilities of each chatbot. This analysis includes assessing the functional correctness of the generated code, its adherence to best practices, and the presence of any code smells. We discuss the implications of our findings, considering both the potential and limitations of AI-generated code in software development.

By employing this methodology, our study not only provides insights into the current state of AI chatbots in code generation but also establishes a framework for future evaluations, contributing to the ongoing improvement and application of AI technologies in the field of software development.

\subsection{Limitations}
This research is centered around a detailed examination and comparison of the quality of code generated by ChatGPT's stable GPT-3.5 version and Google Gemini, with an emphasis on Python code in our initial analysis and extending to Java in a subsequent, hands-on example. The choice of GPT-3.5, a free version of ChatGPT, is deliberate, underpinning the assumption that it is more widely used due to its accessibility. Our assessment criteria are focused on the functionality, maintainability, and efficiency of the code. It is important to clarify that this study does not investigate the intricate technical details of the chatbots’ architecture, such as how they are trained, their use of reinforcement learning algorithms, or the structure of their neural networks. Moreover, we recognize that ChatGPT and Google Gemini are dynamic technologies, subject to ongoing updates and improvements. Consequently, some of the current limitations we discuss, such as ChatGPT's lack of real-time data access and the token restrictions in API calls, may be overcome in future iterations of these systems.

\section{Related Work}
In the following we present related work as well as similar studies in regard of LLMs.

\subsection{OpenAI ChatGPT}
OpenAI launched ChatGPT, a chatbot based on the GPT-3.5 series, on November 30, 2022, using Reinforcement Learning from Human Feedback (RLHF). Quickly attracting 100 million users in just two months, ChatGPT stands out for its ability to deliver detailed responses across various styles to text inquiries~\cite{dwivedi2023so}. Trained on a vast text dataset, this natural language processing model supports conversational interactions, adeptly handling questions and follow-ups through its user-friendly interface~\cite{lecler2023revolutionizing}.

\subsection{Google Gemini}
Google Gemini, introduced by Google on March 21, 2023 as Bard, operates as an AI chatbot powered by the Pathways Language Model 2 (PaLM 2), which boasts enhancements in mathematical, reasoning, and coding abilities. Unlike typical chatbots, Gemini allows for image inputs in interactions, giving it the capacity to analyze visuals and address queries related to them~\cite{hsiao2023s}. Google Gemini integrates real-time data from a variety of Google services, such as Maps, Flights, Hotels, and YouTube, to offer detailed answers in a unified manner. Users can also link their Google Workspace accounts, permitting Gemini to pull and provide information from Docs, Drive, and Gmail. Google Gemini facilitates exporting its responses to several Google applications, including Gmail and Docs, and credits its information sources by listing reference links with its replies.

\subsection{Github Copilot}
GitHub Copilot, an AI-based coding assistant described as an "AI pair programmer," utilizes the OpenAI Codex model for intelligent code completions, drawing on extensive coding and language data. Pearce et al.~\cite{pearce2022asleep} evaluated Copilot's security by testing it against scenarios based on the 2021's top 25 most dangerous software weaknesses (Common Weakness Enumerations, or CWE). They crafted 89 scenarios, resulting in 1,689 programs, and found that about 40\% of the generated code had security vulnerabilities. This indicates the need for developers to be cautious of potential security risks when using Copilot and suggests combining it with security-focused tools to lessen vulnerability risks. Also other research~\cite{asare2023github} comes to similar results. 

\subsection{HumanEval Dataset}
The HumanEval dataset is crafted to test the functional correctness of code produced by Large Language Models (LLMs). It contains 164 intricate programming challenges, each detailed with task ID, prompt, canonical solution, test cases, and entry point. An example from the dataset (HumanEval/12) demonstrates how prompts are used to generate code solutions by chatbots, which are then evaluated against specific test cases and a predefined canonical solution. To ensure the tasks are not part of LLMs' training datasets, all problems in HumanEval are manually created. The dataset covers a wide range of categories, including algorithms (like binary search and sorting), mathematical functions (such as calculating the greatest common divisor or identifying prime numbers), and string functions (for example, finding the longest string in a list or determining string length)~\cite{zheng2023codegeex}.

\subsection{ClassEval Dataset}
The ClassEval dataset is introduced as a more challenging framework to evaluate the code generation capabilities of ChatGPT, Google Gemini, and other LLMs featuring 100 classes and 412 methods designed for complex coding scenarios, especially those involving classes with interdependent methods. Unlike the HumanEval dataset, which focuses on simpler, independent tasks, ClassEval aims to mirror real-world software development challenges with multiple methods within a class relying on one another. This setup addresses limitations identified in the HumanEval approach, such as the brevity of generated code and the lack of interdependence among tasks. ClassEval includes detailed attributes for each class, such as task ID, code skeleton, tests, solution code, and method dependencies, to facilitate this advanced assessment. Examples from ClassEval include creating a comprehensive HR Management System class and other real-world applications like banking systems, book management databases, and chat systems, emphasizing the practical and intricate nature of software development~\cite{du2023classeval}.
Table \ref{table:comparison_datasets} shows a comparison between the two datasets. 
\begin{table}[h!]
\centering
\begin{tabular}{|p{3cm}|p{5cm}|p{5cm}|}
\hline
\textbf{Feature} & \textbf{HumanEval dataset} & \textbf{ClassEval dataset} \\
\hline
\textbf{Purpose} & To assess the performance of LLMs on simple code units such as functions & To assess the performance of LLMs on complex code units such as classes and methods \\
\hline
\textbf{Focus} & Single and independent Python functions & Complex classes and interdependent Python methods \\
\hline
\textbf{Code Complexity} & Simple code & Complex code \\
\hline
\textbf{Number of Problems} & 164 functions & 100 classes, including 412 methods with an average of 33.1 test cases per class \\
\hline
\textbf{Data Source} & Manually created by human users & Collected from available datasets \\
\hline
\textbf{Supported Languages} & Python & Python \\
\hline
\textbf{Source} & HumanEval & ClassEval \\
\hline
\end{tabular}
\caption{Comparison between HumanEval and ClassEval datasets.}
\label{table:comparison_datasets}
\end{table}

\subsection{Code Quality Metrics}
Understanding and measuring the quality of software code is crucial in developing reliable, maintainable, and efficient software systems. Code quality metrics are quantitative measures that provide insights into the characteristics and performance of code. These metrics are essential for identifying potential areas for improvement and ensuring that the codebase remains robust over time. Among the various aspects of code quality, code smell is a particularly critical concept that refers to any symptom in the source code which may indicate deeper problems.

\paragraph{Code Complexity}~\cite{nunez2017source} metrics measure how complicated a piece of code is. High complexity can make code difficult to understand, test, and maintain. Cyclomatic complexity, which counts the number of linearly independent paths through a program's source code, is a commonly used metric. Lower values are preferable, as they indicate simpler, more comprehensible code.

\paragraph{Code Smells}~\cite{santos2018systematic} are patterns in the code that indicate possible violations of fundamental design principles. They are not bugs—code with smells may function correctly—but they suggest weaknesses that might slow down development or increase the risk of bugs or failures in the future. Common examples include:

\begin{itemize}
    \item Large Class: A class that has grown too large, often by taking on too many responsibilities.
    \item Long Method: A single method that is too long, making it hard to understand.
    \item Duplicate Code: Identical or very similar code exists in more than one location.
    \item Feature Envy: A method that seems more interested in a class other than the one it actually belongs to.
\end{itemize}
Addressing code smells typically involves refactoring, the process of restructuring existing code without changing its external behavior. Refactoring aims to improve the code's internal structure and readability, making it easier to maintain and extend.

\section{Evaluation}
Following the completion of our data collection phase, we now proceed with the evaluation segment of our study. We begin this critical phase by systematically examining the outputs produced by ChatGPT from both the HumanEval and ClassEval datasets. This analysis involves taking the generated source code and validating it with the predefined tests from the datasets. For our evaluation we check first the compilation rate before running the predefined tests. Since we use python the compilation test is checking syntax, names and imports of the code. 

\subsection{Compilation Test}
The results of the compilation test for code generated by ChatGPT are shown in table \ref{table:error_types}, whereas table \ref{table:google_bard_errors} shows the compilation test results for Google Gemini.

\begin{table}[ht]
\centering
\caption{Number of Error Types in ChatGPT solutions for HumanEval and ClassEval datasets}
\label{table:error_types}
\begin{tabular}{|p{4cm}|p{3.5cm}|p{3.5cm}|}
\hline
\textbf{Errors in ChatGPT solutions} & \textbf{HumanEval}  \ \ \ \ \ \ \ \ \ \ \ \ \ \ \ \ (164 tests) & \textbf{ClassEval}  \ \ \ \ \ \ \ \ \ \ \ \ \ \ \ \ \ \ \ \ (100 tests)  \\
\hline
Missing library import & 10 & 5 \\
\hline
Name errors & 3 & 2 \\
\hline
Syntax errors & 1 & 1 \\
\hline
Incomplete code & 0 & 0 \\
\hline
\end{tabular}
\end{table}

\begin{table}[ht]
\centering
\caption{Number of Error Types in Google Gemini Solutions for HumanEval and ClassEval Datasets}
\label{table:google_bard_errors}
\begin{tabular}{|p{4cm}|p{3.5cm}|p{3.5cm}|}
\hline
\textbf{Errors in Google Gemini's solutions compared to overall number of errors} & \textbf{HumanEval}  \ \ \ \ \ \ \ \ \ \ \ \ \ \ \ \ (164 tests) & \textbf{ClassEval} \ \ \ \ \ \ \ \ \ \ \ \ \ \ \ \ \ \ \ \ \ \ (100 tests) \\
\hline
Missing library import & 2 & 4 \\

\hline
Name errors & 3 & 4 \\
\hline
Syntax errors & 3 & 6 \\
\hline
Incomplete code & 0 & 16 \\
\hline
\end{tabular}
\end{table}

In our analysis, it's evident that the majority of the code generated by both ChatGPT and Google Gemini is either immediately compilable or nearly so. Specifically, with ChatGPT, the predominant issue encountered involves missing library imports. This type of error is relatively minor and can be quickly remedied by the programmer or, in many cases, automatically by the Integrated Development Environment (IDE). On the other hand, the most frequent error observed with Google Gemini is the generation of incomplete code. This often occurs when the output exceeds the maximum token limit, resulting in truncated code snippets. While compilation errors such as these and missing imports are less of a concern compared to semantic errors, due to their detectability through automated means, they highlight a need for vigilance~\cite{tan2014bug}. Semantic errors, which pertain to the logic and functionality of the code, are inherently more challenging to identify and rectify as they require a deeper understanding of the code's intended behavior. This distinction underlines the importance of a thorough review and testing process to ensure that code, while compilable, also functions as intended without underlying logical flaws.

\section{Semantic Test}
\begin{figure}[h!]
    \centering
    \includegraphics[width=\textwidth]{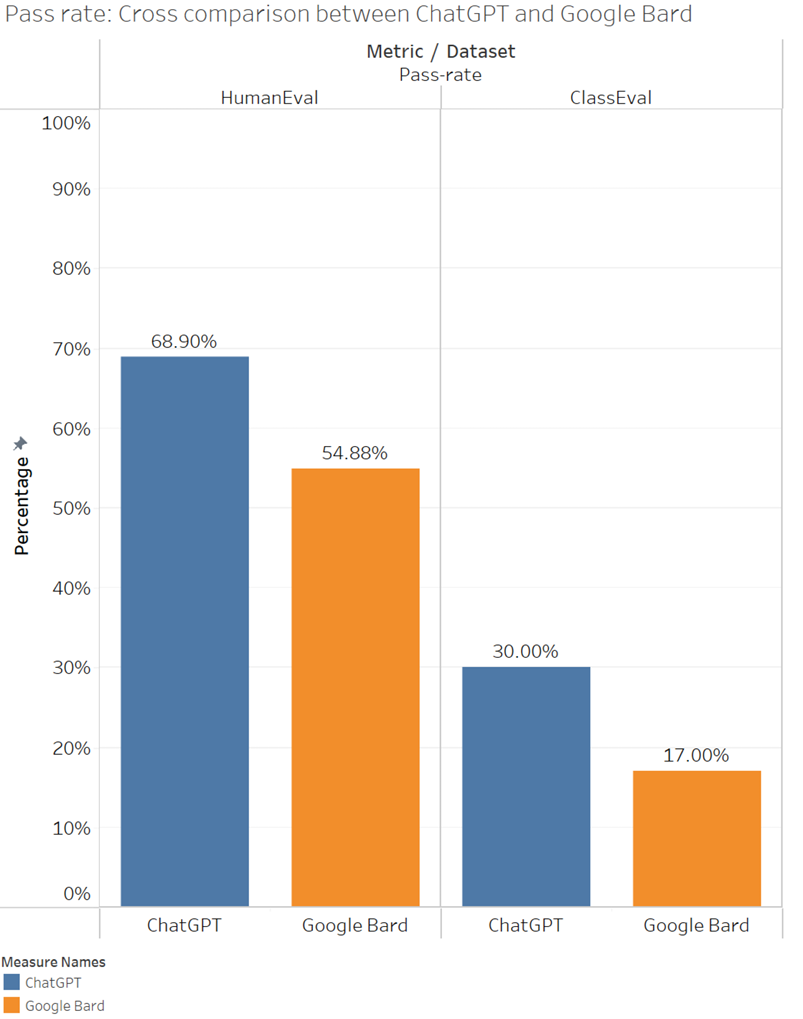}
    \caption{Comparison of the results of the functional tests between ChatGPT and Gemini (former Bard).}
    \label{fig:passing_rate}
\end{figure}

While in many cases of our test scenario the compilation test was passed the results for the semantic test are far more diverse. If the program is working correctly (correct semantic) is tested by functional tests provided by the datasets. The results are shown in Figure \ref{fig:passing_rate}. 

The analysis of the pass rates for ChatGPT and Google Gemini across the HumanEval and ClassEval datasets indicates significant disparities in performance. ChatGPT demonstrated a higher pass rate, with 68.9\% in HumanEval and 30\% in ClassEval, compared to Google Gemini's respective pass rates of 54.878\% and 17\%. This suggests that, in terms of functional correctness, ChatGPT generally delivers better results across both datasets.

However, it's important to contextualize these findings within the realm of practical coding applications. Despite ChatGPT's relatively higher success rates, the results from both AI models are not wholly satisfactory for coding purposes. This is particularly concerning when considering the prevalence of semantic errors, which are inherently more insidious than their syntactic counterparts. Semantic errors, because they pertain to the logic or meaning behind the code, may not be immediately apparent during compilation or initial testing phases. This type of error can lead to incorrect program behavior or output, which might go unnoticed until later stages of development or, worse, deployment. Such issues underscore the need for caution when relying on AI-generated code, highlighting the essential role of thorough code review and testing processes to identify and rectify potentially elusive semantic inaccuracies.

\subsection{Practical Implementation Test}
In an endeavor to further understand the practical implications of utilizing AI-driven coding tools like ChatGPT and Google Gemini in real-world programming scenarios, a secondary experiment was conducted. This test involved software developers tasked with creating a Java program designed to manage a card collection. The program was required to consist of several classes, offering a more intricate challenge that would test both the productivity and the code quality achievable with the assistance of these AI models.

The objective of this test was twofold: to assess the impact of ChatGPT and Google Gemini on speeding up the development process and to evaluate the resulting code's quality. While both AI tools demonstrated a capability to expedite coding tasks significantly, their use was also associated with the introduction of various code smells and minor issues (56 i.e. 72 code smells in 330 lines code for ChatGPT and Google Gemini). 

Moreover, the experiment highlighted a critical aspect of working with AI coding assistants: the necessity of iterative refinement of prompts. Developers often found themselves modifying prompts multiple times to achieve code correctness, indicating that while AI tools can indeed facilitate rapid code generation, they do not always grasp the nuances of the task at hand on the first attempt. Additionally, there was a considerable reliance on the developers' expertise to rectify functionalities and ensure the code met the required standards.

This hands-on test sheds light on the practical benefits and limitations of leveraging AI for software development. While ChatGPT and Google Gemini can undeniably enhance productivity by generating code more quickly than a human programmer might, their propensity to produce code smells and the frequent need for prompt adjustments or manual fixes by the programmer underscore the importance of human oversight. Such findings reinforce the notion that while AI models are powerful tools for accelerating development, they currently serve best as assistants rather than replacements for skilled developers, especially in complex coding projects that demand high standards of code quality and maintainability.

\subsection{Summary}

In our evaluation, we specifically focused on the free versions of ChatGPT and Google Gemini. While it's conceivable that their premium counterparts may offer enhanced performance, including higher accuracy or additional features, the free versions remain highly relevant for a broader audience. This relevance is rooted in the accessibility of these models; they are readily available to a wider range of users, including individual developers, small startups, and educational institutions that might not have the resources to invest in paid subscriptions. Consequently, it's reasonable to assume that a significant portion of programmers and developers engaging with these AI models are utilizing the free versions. This assumption underpins our analysis and underscores the importance of assessing the free versions' capabilities. By examining these accessible versions, our study aims to provide insights that are directly applicable to the majority of users who, driven by cost considerations or trial purposes, opt for the free models. Understanding the performance and limitations of these versions is crucial, as they are likely the primary, if not the only, interaction many users will have with AI-driven coding assistance tools.

\section{Conclusion}
This paper has explored the capabilities of AI models, specifically ChatGPT and Google Gemini, in the context of software development, focusing on their ability to generate programming code. Our findings confirm that these models can significantly accelerate coding tasks, thereby enhancing productivity. This advantage aligns well with the increasing demand for faster development cycles in the tech industry~\cite{asprion2023agile}. However, akin to code crafted by human developers, the output from these AI models is not exempt from flaws. It necessitates rigorous testing~\cite{scherb2023divide} and cautious utilization to ensure the reliability and maintainability of the software produced, especially when used in critical areas~\cite{scherb2023cymed}. Therefore, the LLMs are not releasing the software developers from learning about code quality~\cite{keuning2017code} and security~\cite{scherb2023cyber}\cite{scherb2023serious}.

The presence of code smells, semantic errors, and other minor issues in the code generated by ChatGPT and Google Gemini highlights the critical role of thorough testing protocols. These AI-driven tools, while powerful, are not infallible; they require human oversight to correct inaccuracies and refine functionalities. As such, these models should be viewed as valuable assistants in the coding process, capable of streamlining development but not replacing the nuanced understanding and decision-making capabilities of experienced programmers.

\section{Future Work}
Looking forward, there are several avenues for further research that could enrich our understanding of AI's role in software development. Firstly, an evaluation of the premium versions of ChatGPT and Google Gemini would be insightful. Such an analysis could reveal whether the enhanced features and capabilities of the paid models address some of the limitations identified with their free counterparts, potentially offering even greater benefits in terms of code quality and developer productivity.

Additionally, conducting a study with companies that are integrating these LLMs into their software development processes could provide valuable real-world insights. Examining how businesses utilize AI coding assistants on a day-to-day basis, and the impact this has on project timelines, code quality, and overall productivity, would offer a comprehensive view of the practical applications and challenges of these technologies in the industry.

A further approach can be to let LLMs write their own Unit tests which was quite successful in current preprints~\cite{siddiq2024using}.
%
%
%
 \bibliographystyle{splncs04}
 \bibliography{mybibliography}

\end{document}